\documentclass[global,twocolumn]{svjour}
\usepackage{graphics}
\usepackage{amsmath}
\journalname{Applied Physics B}
\begin{document}
\title{Paramagnetic Faraday rotation with spin-polarized ytterbium atoms}
\author{M. Takeuchi\inst{1} \and T. Takano\inst{1} \and S. Ichihara\inst{1} \and Y. Takasu\inst{2} \and M. Kumakura\inst{1}\inst{3}\inst{4} \and T. Yabuzaki\inst{5} \and and Y. Takahashi\inst{1}\inst{4} \thanks{Fax:+81-75-753-3769, E-mail:yitk@scphys.kyoto-u.ac.jp}
}
\institute{
 Department of Physics, Graduate School of Science, Kyoto University, Kyoto 606-8502, Japan
 \and
 Department of Electronic Science and Engineering, Graduate School of Engineering, Kyoto University, Kyoto 615-8510, Japan
 \and
 PREST, JST, 4-1-8 Honcho Kawaguchi, Saitama 332-0012, Japan
 \and
 CREST, JST, 4-1-8 Honcho Kawaguchi, Saitama 332-0012, Japan
 \and
 Faculty of Information Science and Arts, Osaka Electro-Communication University, Osaka 572-8530, Japan
}
\date{Received: / Revised version:}
\maketitle
\begin{abstract}
We report observation of the paramagnetic Faraday rotation 
 of spin-polarized ytterbium (Yb) atoms.
As the atomic samples,
 we used 
 an atomic beam,
 released atoms from a magneto-optical trap (MOT),
 and trapped atoms in a far-off-resonant trap (FORT).
Since Yb is diamagnetic and includes a spin-1/2 isotope,
 it is an ideal sample for the spin physics, such as 
 quantum non-demolition measurement of spin (spin QND),
 for example.
From the results of the rotation angle,
 we confirmed that the atoms were almost perfectly polarized.

\end{abstract}

\section{Introduction}

Faraday rotation is the polarization rotation of
 linearly-polarized light
 due to the circular birefringence of the medium.
When the refractive index of the medium is $n_\pm$ for
 circularly polarized light $\sigma_\pm$, respectively,
 the Faraday rotation angle $\phi$ becomes
\begin{align}
 \phi=\frac{\omega L}{2c}(n_+-n_-),
\end{align}
where $\omega$ is the angular frequency of the probe light,
 $L$ is the length of the medium,
 and $c$ is the light velocity \cite{budker02}.
Especially, the paramagnetic Faraday rotation
 is the powerful method to probe the spin state
 of an atomic ensemble \cite{happer72}.
The rotation angle for an atomic ensemble
 via the paramagnetic Faraday rotation can be written as
\begin{align}
 \phi=\frac{\alpha t_1}{2}S_z,\label{PFR}
\end{align}
 where $\alpha$ is a real constant,
 $t_1$ is the intreaction time,
 and $S_z$ is the total spin component in the probe region
 parallel to the propagation direction of light \cite{takahashi99}.
\begin{figure}[h]
 \begin{center}
 \includegraphics{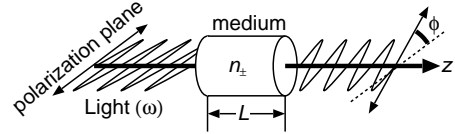}
 \caption{
 Faraday rotation is the polarization rotation of
 linearly-polarized light
 due to the circular birefringence of the medium ($n_\pm$).
 In the case of the paramagnetic Faraday rotation,
 the rotation angle is proportional to the total spin 
 component parallel to the propagation direction of light ($S_z$).
 }
 \end{center}
\end{figure}

Recently, there has been a renewal of interest in 
 the paramagnetic Faraday rotation
 as quantum non-demolition measurement of spin
 (spin QND) \cite{takahashi99,kuzmich98,kuzmich99}.
Spin QND is not only a quantum measurement,
 but also has a wide variety of the applications.
For example, spin squeezing, entanglement of two macroscopic objects, 
 and quantum memory for light have been demonstrated
 \cite{kuzmich00,julsgaard01,geremia04,julsgaard04},
 and the reversible quantum measurement has been proposed
 \cite{terashima05}.
These applications of spin QND will be some breakthroughs
 in not only quantum information processings
  but also precision measurements \cite{takahashi97}.

The ytterbium (Yb) atoms has many merits for spin QND.
(i) Since the electronic ground state of Yb is diamagnetic
 ($\mathrm{{}^1S_0}$),
 the magnetic moment has its origin only in nuclear spin,
 which is three order of magnitude smaller than the paramagnetic atoms.
It is experimentally easier to suppress 
 the fluctuation of the precessions
 of the spin than that by use of the paramagnetic atoms like Ref.
 \cite{kuzmich99,kuzmich00,julsgaard01,geremia04,julsgaard04,isayama99}.
Therefore, the longer coherence time will be expected.
Moreover, the spin polarization is also easier
since the optical pumping process is faster than the Larmor precession.
(ii) Yb atoms include a spin-1/2 isotope ($\mathrm{{}^{171}Yb}$).
Since the interaction between spin-1/2 particle and
 the external electro-magnetic field
 is described by scalar and vetor polarizabilities alone,
 instead of more general tensor polarizability
 \cite{happer72},
 it can be said that $\mathrm{{}^{171}Yb}$ is one of the most
 ideal sample for the spin physics.
(iii) Yb atoms include also spin-0 isotopes.
By comparing the signal from spin-1/2 isotope
 with the signals from spin-0 isotopes,
 the calibrations of the experimental system are possible.
In Table \ref{spin}, we summarize the isotopes and the aboundance
 of ytterbium.
\begin{table}[h]
\caption{
 Mass number and natural abandance of stable ytterbium categolized
 by the nuclear spin.
}
\label{spin}
\begin{center}
\begin{tabular}{rll}
\hline\noalign{\smallskip}
Nuclear Spin ($I$) & Mass Number ($M$) & Abandance  \\
\noalign{\smallskip}\hline\noalign{\smallskip}
0 & 168,170,172,174,176 & 69.5~\% \\
1/2 & 171 & 14.3~\% \\
5/2 & 173 & 16.2~\% \\
\noalign{\smallskip}\hline
\end{tabular}
\end{center}
\end{table}
(iv) The laser-cooling and trapping techniques have already 
 been established
 \cite{honda99,kuwamoto99,honda02,takasu03}.

As far as we know, however,
 the observation of the paramagnetic Faraday rotation
 with spin-polarized Yb atoms has not been reported yet.
In this paper, we report the theoretical estimations
 of the rotation angle, and its observations.
As the samples,
 we used atomic beam,
 released atom from the magneto-optical trap (MOT),
 and trapped atom in far-off-resonant-trap (FORT).
Simultaneously,
 we have almost perfectly polarized 
 $\mathrm{{}^{171}Yb}$ and $\mathrm{{}^{173}Yb}$
 via the optical pumping \cite{happer72}.
By our experimental results,
 Yb will be considerd as one of the best sample for spin QND.

\section{Theory}
In our experiment, $\omega$ is close to the
 resonance frequency of $\mathrm{{}^1S_0\to{}^1P_1}$ transition
 $\omega_0$.
The electric dipole moment $\mu_e$ can be written as
\cite{metcalf99}
\begin{align}
 \mu_e^2=\frac{e^2\sigma_0 \Gamma}{8\pi\alpha_\mathrm{f} \omega_0},
\end{align}
 where $\alpha_\mathrm{f}$ is the fine structure constant,
 $e$ is the charge of an electron,
 $\sigma_0\equiv 6\pi(c/\omega_0)^2$ is 
 the photon-absorption cross section of an atom,
 and $\Gamma$ is the natural full linewidth of the transition
 in angular frequency.
In the following calculations, we assume $\omega/\omega_0\simeq 1$,
 and that the electric field of light $E$,
 or the intensity $I$ is weak,
\begin{align}
 \Omega^2\equiv
 \frac{\mu_e^2 E^2}{\hbar^2}
 =\frac{\Gamma^2I}{2I_s}
 \ll (\omega_0-\omega)^2+(\Gamma/2)^2,
\end{align}
 which corresponds to the elimination
 of the higher-order electric susceptibility.
$\Omega$ is the Rabi frequency, 
 and $I_s$ is the saturation intensity given by
 $I_s=\hbar\omega_0\Gamma/2\sigma_0$ \cite{metcalf99}.
For Yb,
 $\omega_0=2\pi\times 751.5~\mathrm{THz}$,
 $\Gamma=2\pi\times 29~\mathrm{MHz}$,
 $\sigma_0=7.598\times 10^{-14}~\mathrm{m^2}$,
 and $I_s=0.60~\mathrm{mW/mm^2}$.

\subsection{Spin-0 isotopes:
$\mathit{{}^{168}Yb,{}^{170}Yb,{}^{172}Yb,{}^{174}Yb,{}^{176}Yb}$}

\begin{figure}[h]
 \includegraphics{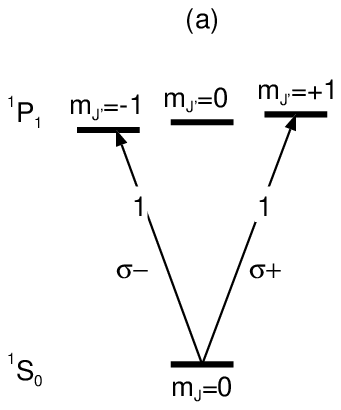}
 \includegraphics{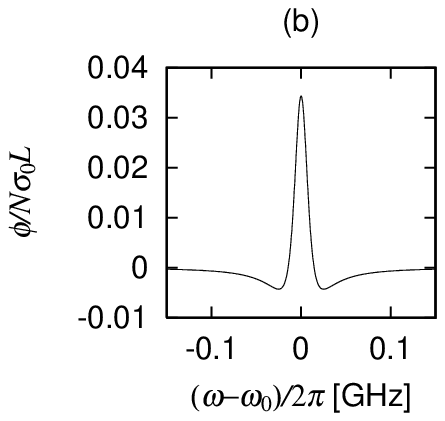}
 \caption{
    (a)Level structure and the transition probabilities
    for $\sigma_\pm$ light of spin-0 isotopes.
    (b)Theoretical calculation of the rotation angle when
    $\omega_{\pm 1}=\omega_0\pm 2\pi\times 1~\mathrm{MHz}$.
 }
 \label{level-0}
\end{figure}

In Fig. \ref{level-0},
 we depict the level structure and the transition probabilities
 for $\sigma_\pm$ light of spin-0 isotopes.
The refractive indices $n_\pm$ become \cite{suter97}
\begin{align}
 n_\pm&=1+\frac{2\pi\alpha_\mathrm{f} c\mu_e^2}{e^2}
 g_{\pm 1}N \label{n_pm-0}
\end{align}
 where $N$ is the number density of the atoms,
$g_{m_{J'}}$ is given by
\begin{align}
 g_{m_{J'}}=\frac{\omega_{m_{J'}}-\omega}
 {(\omega_{m_{J'}}-\omega)^2+(\Gamma/2)^2},
\end{align}
 and is the dispersive function of $\omega$ at the center frequency
 $\omega_{m_{J'}}$
 which is the resonance frequency between the ground state and
 the each excited state $m_{J'}$.
From Eq. (\ref{n_pm-0}), the rotation angle becomes
\begin{align}
 \phi=\frac{\Gamma}{8}(g_{+1}-g_{-1})N\sigma_0L. \label{phi-0}
\end{align}
It should be noted that the rotation angle is proportional to
 $N\sigma_0L$.
From Eq. (\ref{phi-0}) it is obvious that the rotation angle
 vanishes when the excited states are degenerated
 such as $\omega_{+1}=\omega_{-1}$.
This is not the paramagnetic Faraday rotation
 but the diamagnetic Faraday rotation,
 and is useful for magnetic field compensation.
Since the magnetic moment of $\mathrm{{}^1P_1}$ state is as large as 
 one Bohr magneton,
 the stray magnetic field parallel to the propagation of light,
 which leads the Zeeman splitting of the sublevels 
 $m_{J'}=\pm1$ in the $\mathrm{{}^1P_1}$ state,
 can be monitored via this Faraday rotation.
In Fig. \ref{level-0}(b),
 we plot Eq. (\ref{phi-0}) as a function of $\omega$.
It should be noted that the rotation angle rapidly decreases
 at off-resonance.

\subsection{Spin-1/2 isotope:
$\mathit{{}^{171}Yb}$}

\begin{figure}[h]
  \includegraphics{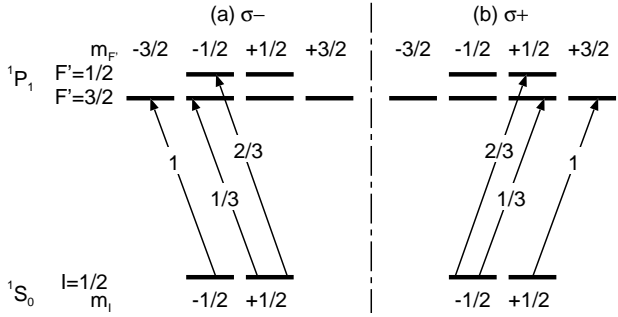}
  \caption{
   Levels and squared transition probabilities of spin-1/2 isotope.
   (a) For $\sigma -$ light.
   (b) For $\sigma +$ light.
  }
  \label{level-1_2}
\end{figure}

\begin{figure}[h]
 \includegraphics{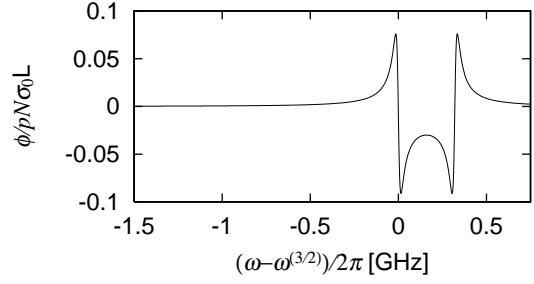}
 \caption{Rotation angle of spin-1/2 isotope.
 Here, we used the value of the hyperfine splitting,
 $\omega^{(1/2)}-\omega^{(3/2)}=2\pi\times 320~\mathrm{MHz}$
 \cite{banerjee03}.
 }
 \label{pfr-1_2}
\end{figure}

To simplify the discussions in the followings,
 we assume that the applied magnetic field is very weak,
 and so the magnetic sublevels are almost degenerate,
 and therefore we do not consider the diamagnetic Faraday rotation
 discussed above.
For this isotope,
 the good quantum number of the $\mathrm{{}^1P_1}$ state is $F'=J'+I$.
To derive the transition probabilities of each transition,
 the additional rule of angular momenta must be considered
 \cite{sakurai95}.
The expansion coeffients
 $\langle J',I;m_{J'},m_I|J',I;F',m_{F'}\rangle$ can be calculated as
\begin{align}
\left|\langle 1,1/2;\pm 1,\pm 1/2|1,1/2;3/2,\pm 3/2\rangle\right|^2
 =&1,\\
\left|\langle 1,1/2;\pm 1,\mp 1/2|1,1/2;3/2,\pm 1/2\rangle\right|^2
 =&\frac{1}{3},\\
\left|\langle 1,1/2;\pm 1,\mp 1/2|1,1/2;1/2,\pm 1/2\rangle\right|^2
 =&\frac{2}{3},
\end{align}
and the other coeffients become zero.
In Fig. \ref{level-1_2}, we depict
 the levels and the probabilities of the each transitions
 for $\sigma_\pm$ light.
The refractive indices become
\begin{align}
 n_\pm =&1+\frac{2\pi\alpha_\mathrm{f} c\mu_e^2}{e^2}
  \times\nonumber\\
  &\left(g^{(3/2)} N_{\pm 1/2}
  +\frac{2}{3}g^{(1/2)} N_{\mp 1/2}
  +\frac{1}{3}g^{(3/2)} N_{\mp 1/2}\right) \label{n_pm-1_2},
\end{align}
 where $g^{(F')}$ is the dispersive function
 at the center frequency $\omega^{(F')}$,
 which is the resonance frequency between the ground and
 each hyperfine excited state $F'$,
 and $N_{\pm 1/2}$ is the number density of atom in the
 ground state $m_I=\pm 1/2$, respectively.
From Eq. (\ref{n_pm-1_2}), the rotation angle becomes
\begin{align}
 \phi=\frac{\Gamma}{12}\left(g^{(3/2)}-g^{(1/2)}\right)
 pN\sigma_0L,
 \label{phi-1_2}
\end{align}
 where $p\equiv(N_{+1/2}-N_{-1/2})/N$ is the spin polarization.
It is obvious that the rotation angle vanishes
 when there is no population difference between the
 ground sublevels such as $p=0$,
 and is proportional to $N\sigma_0 L$
 as for the diamagnetic Faraday rotation of Eq. (\ref{phi-0}).
In. Fig. \ref{pfr-1_2}, we plot Eq. (\ref{phi-1_2})
 as a function of $\omega$.
It should be noted that the rotation angle slowly decreases
 at off-resonance.

Since the total spin component of an atomic ensemble is written as 
 $S_z=pN\pi w^2L/2$, that is, $p$ is proportional to $S_z$,
 it can be said that from Eq. (\ref{phi-1_2})
 the rotation angle $\phi$ always reflects $S_z$ 
 for any detuning,
 where $w$ is the beam waist of the probe light.
In the case of the atoms having complicated level structures,
 the relation of Eq. (\ref{PFR}) is only satisfied
 for large detunings \cite{isayama99}.
The coupling constant $\alpha t_1$ decreases, however,
 for large detunings.
This is one of the reasons why spin-1/2 isotope is 
 the most ideal sample for spin QND.

\subsection{Spin-5/2 isotope:
$\mathit{{}^{173}Yb}$}

\begin{figure}[h]
 \includegraphics{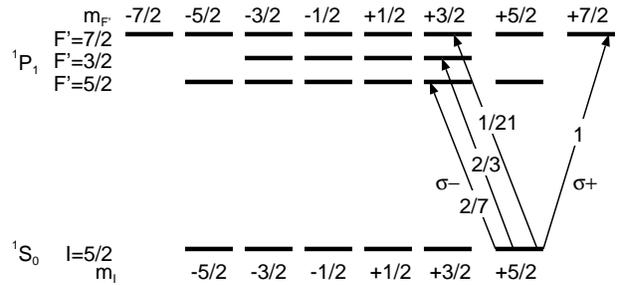}
 \caption{
    Levels and some squared transition probabilities
    for spin-5/2 isotope.
 }
 \label{level-5_2}
\end{figure}

\begin{figure}[h]
 \includegraphics{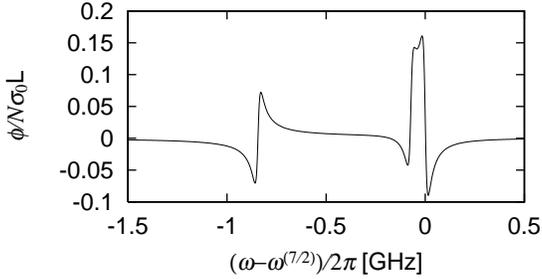}
 \caption{
    Rotation angle of spin-5/2 isotope.
    Here we used the value of the hypefine splitting,
    $\omega^{(3/2)}-\omega^{(7/2)}=-2\pi\times 73$~MHz,
    and $\omega^{(5/2)}-\omega^{(7/2)}=-2\pi\times 844$~MHz
    \cite{banerjee03},
    and we assumed the population is polarized at $m_I=+5/2$ state.
 }
 \label{pfr-5_2}
\end{figure}

The level structure of spin-5/2 isotope is complicated
 as is shown in Fig. \ref{level-5_2}.
To simplify the discussion,
 we consider the case that the atom is polarized
 at $m_I=+5/2$ state in the ground state.
From the transition probabilites, the rotation angle becomes
\begin{align}
  \phi=\frac{\Gamma}{84}
  \left(10g^{(7/2)}-7g^{(3/2)}-6g^{(5/2)}\right)N\sigma_0L,
  \label{phi-5_2}
\end{align}
 where $N$ is the number density of the atom.
In Fig. \ref{pfr-5_2}, we plot Eq. (\ref{phi-5_2}) as a function of $\omega$.
It should be noted that the rotation angle vanishes
 when there is no population difference among the ground sublevels,
 similar to the case of spin-1/2 isotope.

\subsection{Spin polarization}

\begin{figure}[h]
 \begin{center}
 \includegraphics{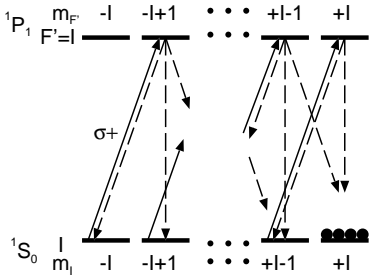}
 \caption{
  Schematic of the optical pumping.
  The atom in $m_I\neq +I$ state absorb $\sigma_+$ light
   (solid lines),
   then emit $\sigma_\pm$ or $\pi$ light (dot lines)
   and fall to a state.
   By repeating the absorptions and the emissions,
   all population finally transfers to $m_I=+I$ state.
 }
 \label{level-pump}
 \end{center}
\end{figure}

As is mentioned above, the rotation angle has non-vanishing value
 when there is some population difference
 among the ground sublevels.
To make the population difference in the ground states,
 we performed the optical pumping by using
 $\mathrm{{}^1S_0}\to\mathrm{{}^1P_1}(F'=I)$ transition with
 circulary polarized light.
In Fig. \ref{level-pump}, we describe the schematic of the
 optical pumping with $\sigma_+$ light.
The atom in $m_I\neq +I$ state absorb $\sigma_+$ light,
 then emit $\sigma_\pm$ or $\pi$ light and fall to a state.
By repeating the absorptions and the emissions,
 all population finally transfers to $m_I=+I$ state.
The details of the optical pumping are written in some references
 \cite{suter97}.

\section{Experiments}

\begin{figure}[h]
 \begin{center}
 \includegraphics{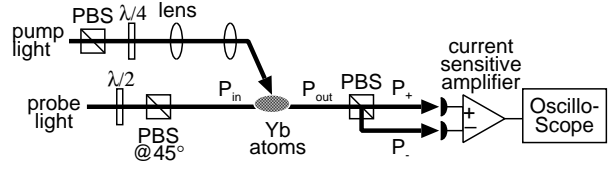}
 \caption{
  Basic setup of the system.
  The paramagnetic Faraday rotation appears
   after the optical pumping with circulary polarized light.  
  The rotation angle is monitored with a polarimeter.
  PBS:polarization beam splitter,
  $\lambda/4$: quater-wave plate,
  $\lambda/2$: half-wave plate.
  $P_\mathrm{in}$, $P_\mathrm{out}$, $P_+$ and $P_-$
  is the power at the each point.
  }
 \label{pump-probe}
 \end{center}
\end{figure}

For our experiments,
 we used the external cavity laser diode (ECLD)
 and injection-locking of laser diode (LD) technique \cite{komori03}.
In Fig. \ref{pump-probe}, we show the basic setup to observe
 the paramagnetic Faraday rotation.
The pump light is magnified with lenses
 so as to cover the atom distribution.
As the monitor of the rotation angle,
 we constructed a polarimeter which consists of two photodiodes (PD)
 and a current sensitive amplifier.

Here, we note the property of the polarimeter.
We define $P_\mathrm{in}$, $P_\mathrm{out}$, $P_+$ and $P_-$
 as the power in front of the atoms,
 behind the atoms,
 in front of the PD of the sense $+$,
 and in front of the PD of the sense $-$, 
 respectively.
The output of the polarimeter is proportional to
 $P_+-P_-=P_\mathrm{out}\sin(2\phi)$.
$P_\mathrm{out}$ can be measured by the relation
 $P_\mathrm{out}=P_++P_-$.
We also define the optical depth as
 $-\mathrm{ln}(P_\mathrm{out}/P_\mathrm{in})$.
The optical depth for resonance is useful since
 it reflects $N\sigma_0L$,
 which is the proportional factor of the Faraday rotation
 as is shown in Eq. (\ref{phi-0}), Eq. (\ref{phi-1_2}),
 and Eq .(\ref{phi-5_2}).
In the case of spin-nonzero isotopes,
 we must take into consideration the relative transition probabilities
 among the excited hyperfine states and the ground state to deduce $N\sigma_0L$.
In Table \ref{probabilities},
 we show the factors.
\begin{table}[h]
\caption{
 Relative transition probabilities among the hyperfine sublevels
 for $\pi$ polarized light.
}
\label{probabilities}
\begin{center}
\begin{tabular}{cc|cc}
\hline\noalign{\smallskip}
$M(F')$ & Probability & $M(F')$ & Probability\\
\noalign{\smallskip}\hline\noalign{\smallskip}
171(1/2) & 1/3 & 173(7/2) & 4/9\\
171(3/2) & 2/3 & 173(3/2) & 2/9\\
         &     & 173(5/2) & 1/3\\
\noalign{\smallskip}\hline
\end{tabular}
\end{center}
\end{table}

\subsection{Polarization spectroscopy with atomic beam}

Firstly, we observed the paramagnetic Faraday rotation
 with atomic beam.
By use of the atomic beam,
 we could continuously observe the signal.
The number density of the atomic beam was stable,
 and less insensitive for the photon scattering.
By pumping with circulary polarized light
 of an appropriate frequency,
 we could isotope-selectively observe
  the paramagnetic Faraday rotation.

\begin{figure}[h]
 \begin{center}
 \includegraphics{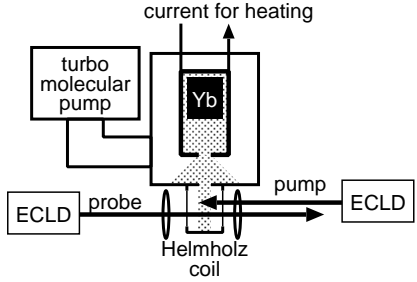}
 \caption{
  Atomic beam system and experimental setup
   for the polarization spectroscopy.
  The atomic oven was heated by electric current
   to obtain the high-temperature $500~\mathrm{{}^\circ C}$.
  The atomic beam was collimated by an apparture of the diameter
   $L=5~\mathrm{mm}$.
  They were in a chamber vacuuated by a turbo molecular pump.
  A Helmholtz coil was set around the atomic beam 
   so as to cancel the stray magnetic field parallel to the probe
   light.
 }
 \label{atomicbeam}
 \end{center}
\end{figure}

The atomic beam was generated from metallic Yb sample
 in an atomic oven of about $500\mathrm{~{}^\circ C}$.
The oven was in a vacuum chamber of the pressure
 $3\times10^{-4}~\mathrm{Pa}$.
A Helmholtz coil was set around the atomic beam
 so as to cancel the stray magnetic field of about
 $1\times10^{-4}~\mathrm{T}$
 parallel to the probe light.
In Fig. \ref{atomicbeam},
 we show the apparatus.

\begin{figure}[h]
 \begin{center}
 \includegraphics{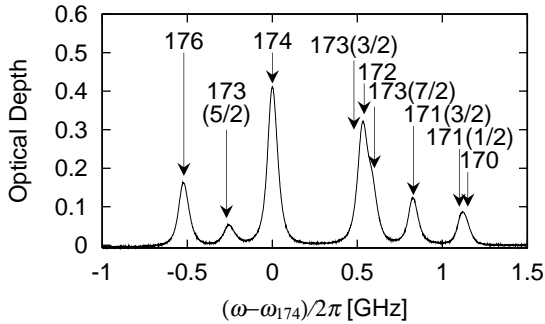}
 \caption{
  Absorption spectrum of the atomic beam.
  It shows that the atomic beam was so well collimated
   that the linewidth caused by the transverse 
  Doppler broadening is $\Gamma^*=2\pi\times 57~\mathrm{MHz}$,
  about twice of the natural linewidth $\Gamma$.
  $\omega_{174}$ is the resonance frequency of $\mathrm{{}^{174}Yb}$,
  which is the most naturally aboundant isotope (31.8~\%).
  The number in the parentheses represent hyperfine quantum number.
 }
 \label{absorption}
 \end{center}
\end{figure}

In Fig. \ref{absorption},
 we show the absorption spectrum with the probe light.
It shows that the atomic beam was so well collimated
 that the linewidth caused by the transverse 
 Doppler broadening is $\Gamma^*=2\pi\times57~\mathrm{MHz}$,
 about twice of the natural linewidth $\Gamma$,
 obtained from the fitting with 
  the isotope shifts and the hyperfine splittings \cite{banerjee03},
  and the relative transition probabilities for $\pi$ polarized light
  among the hyperfine sublevels (Table \ref{probabilities}).
In the fitting,
 we approximated the Doppler broadeneng $\Gamma^*$
  as the natural line width $\Gamma$,
  known as $T_2^*\to T_2$ approximation.
From Fig. \ref{absorption}, we obtained
 $N\sigma_0L=0.18$ for $\mathrm{{}^{171}Yb}$ and
 $N\sigma_0L=0.21$ for $\mathrm{{}^{173}Yb}$.

\begin{figure}[h]
 \begin{center}
 \includegraphics{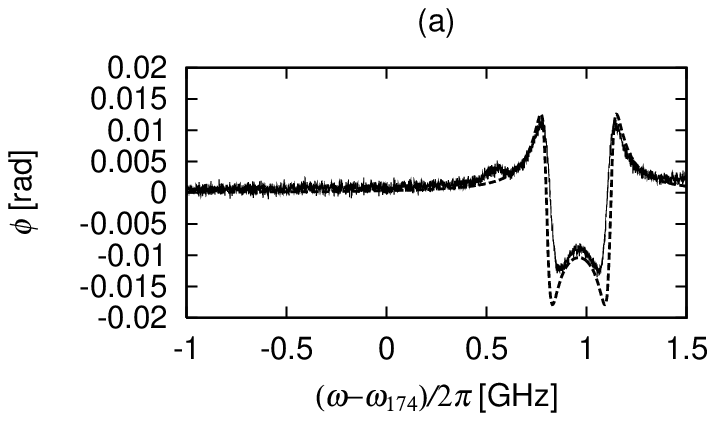}
 \includegraphics{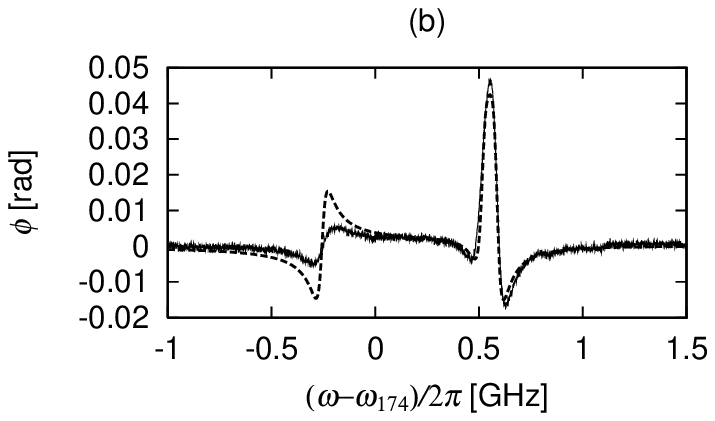}
 \caption{
  Results of the Faraday rotation (solid lines),
   and the theoretial curves assuming
    the perfect polarization (dot lines).
  (a) The frequency of the pumping light was tuned at
   $171(1/2)$ resonance.
  (b) The frequency of the pumping light was tuned at
   $173(5/2)$ resonance.
 }
 \label{polspec}
 \end{center}
\end{figure}

In Fig. \ref{polspec},
 we show the Faraday rotation with the optical pumping.
Simultaneously, we show the theoretical curves assuming
 the $T_2^*\to T_2$ approximations and
 the perfect polarizations by the optical pumping.
In Fig. \ref{polspec} (a) and (b) of the theoretical curve,
 we made correction for the resonance frequency
 $\omega^{(3/2)}$ and $\omega^{(7/2)}$, respectively,
 for the imperfect linearlity for the frequency sweep
 of our ECLD.
As is shown in Fig. \ref{polspec},
 the rotation angle except for the near resonance well agrees
 with the theoretical estimation,
 which indicates that the optical pumping was ideally performed.

The rotation angles were small at some near-resonant frequencies
 compared with the theoretical values.
We think that this is due to the depolarization by the probe light.
The photon scattering rate of two-level system
 can be written as \cite{suter97}
\begin{align}
r=\frac{\Gamma}{4}
\frac{\Omega^2}{(\omega_0-\omega)^2+(\Gamma/2)^2+(\Omega/2)^2}.
 \label{r}
\end{align}
The transit time of the probe region for an atom $T$
 can be estimated as $T\sim 2w/v= 0.9~\mathrm{\mu s}$,
 where $w$ is the probe beam waist measured as $w=0.14~\mathrm{mm}$
 and $v$ is the squared-average longitudinal velocity
 of the atomic beam estimated as $v=0.3~\mathrm{km/s}$
 from the oven temperature.
Since the intensity of the probe light at the atomic beam region
 was $0.55~\mathrm{mW/mm^2}$,
 the photon absorption rate at the peak of the dispersive function
 $\omega_0-\omega=\Gamma^*/2$ becomes
 $r\sim4\times10^7~\mathrm{s^{-1}}$.
Therefore the scattering counts becomes $rT\sim 4\times 10^1$,
 which is large enough to decrease the population differences.
Strictly speaking,  Eq. (\ref{r}) is inappropriate to discuss
 the photon scattering rate at near-resonance,
 because $\mathrm{{}^{171}Yb}$ and $\mathrm{{}^{173}Yb}$
 are not the two-level system.
However, we emphasize that the rotation angle
 except for the near resonance well agrees with
 the theoretical estimation.

As Eq. (\ref{phi-0}) indicates,
 the stray magnetic field parallel to the probe light
 leads to the diamagnetic Faraday rotation.
In Fig. \ref{polspec},
 the rotation angle well vanished near the resonance frequency
 of $\mathrm{{}^{170}Yb,{}^{172}Yb,{}^{174}Yb,{}^{176}Yb}$,
 which indicates that the stray magnetic fields
 were well suppressed by the Helmholtz coil,
 otherwise we could in fact observe the Faraday rotation
 like Fig. \ref{level-0}(b).
We emphasize that the magnetic field is not necessary
 for spin polarization because the optical pumping process
 is faster than the Larmor precession.
This is the reason why Yb can be highly polarized easily.

\subsection{Ballistically expanding cold $\mathit{{}^{171}Yb}$ atoms released from MOT}

Secondly, we observed the paramagnetic Faraday rotation
 of the ballistically expanding cold atom released from MOT.
By use of cold atom,
 it became possible to probe an atomic ensemble
 for a long time,
 which was impossible for the atomic beam
 due to the longitudinal velocity.

\begin{figure}[h]
 \begin{center}
 \includegraphics{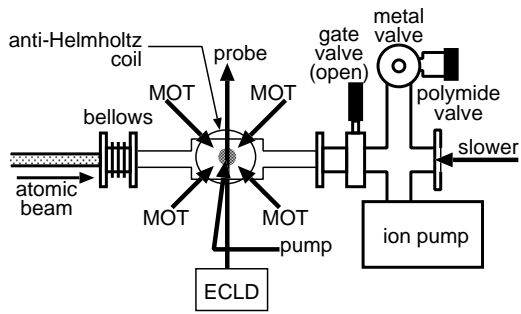}
 \caption{
 Experimental setup.
 By using a glass cell at the MOT chamber,
  we could obtain good optical access to atoms.
 The background gas pressure in the glass cell was about
  $5\times10^{-6}~\mathrm{Pa}$. 
 The magnetic gradient created by the anti-Helmholz coil 
  was $4.3\times10^{-4}~\mathrm{T/mm}$ along the axis.
 }
 \label{mot}
 \end{center}
\end{figure}

\begin{figure}[h]
 \begin{center}
 \includegraphics{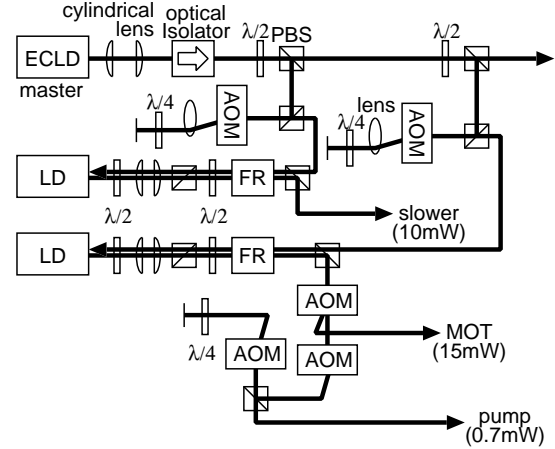}
 \caption{
  Schematic of the light source
   for the MOT and optical pumping of
   $\mathrm{{}^{171}Yb}$.
   FR:Faraday rotator, AOM:acousto-optic modulator.
   The typical powers are also shown.
 }
 \label{injectionlock}
 \end{center}
\end{figure}

\begin{figure}[h]
 \includegraphics{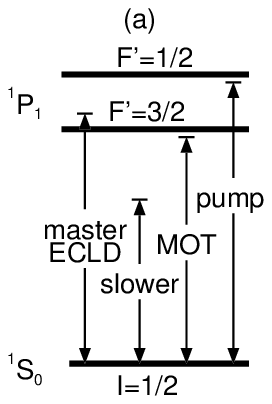}
 ~~~~~\includegraphics{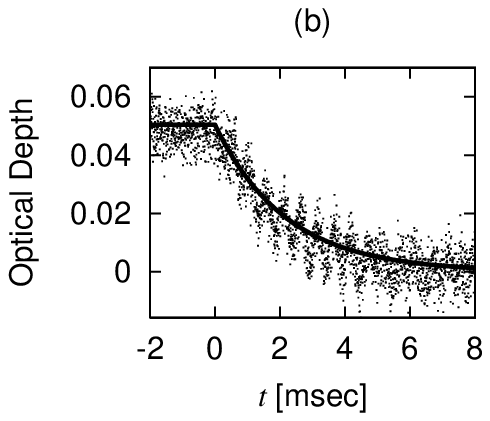}
 \caption{
  (a) Each frequency of light.
  The frequency of the master ECLD
   was tuned at the red side of $\omega^{(3/2)}$,
   that of the slower light was at -216~MHz from $\omega^{(3/2)}$,
   that of MOT light was at -26~MHz from $\omega^{(3/2)}$,
   and that of the pumping light was close to $\omega^{(1/2)}$.
  (b) The optical depth of the MOT measured by the probe light whose
   frequency was tuned at $\omega^{(3/2)}$.
  The dot is the experimental value,
   and the line is the fitting curve.
 }
 \label{level-mot}
\end{figure}

In Fig. \ref{mot},
 we show the experimental setup.
By using a glass cell ($50\times50\times100~\mathrm{mm}$)
 at the MOT chamber,
 we could obtain good optical access to atoms in the MOT.
The background gas pressure in the glass cell was about
 $5\times 10^{-6}~\mathrm{Pa}$.
The magnetic gradient created by the anti-Helmholz coil 
 was $4.3\times10^{-4}~\mathrm{T/mm}$ along the axis.
In Fig. \ref{injectionlock},
 we show the system of the light source.
By using the injection-lock techniques \cite{komori03},
 the frequency stabilizations of the two LDs were 
 experimentally simplified.
In Fig. \ref{level-mot} (a), we show their frequencies.
The frequency of the master ECLD
 was tuned at the red side of $\omega^{(3/2)}$,
 that of the slower light was at -216~MHz from $\omega^{(3/2)}$,
 that of MOT light was at -26~MHz from $\omega^{(3/2)}$,
 and that of the pumping light was close to $\omega^{(1/2)}$.

The temperature of the trapped atom was
 typically 4~mK estimated from the release and recapture method
 with the MOT beam diameter of 10~mm
 and the release time 7.5~ms.
The trap lifetime was 0.3~s,
 measured by the decay of the fluorescence
 after switching off the slower beam.
The diameter of the trapped atom cloud was 
 $L\sim2~\mathrm{mm}$ roughly estimated by a CCD camera.
The trapped atom number was
 typically estimated as $7\times 10^6$
 from the intensity of the fluorescence.
It should be noted that we also succeeded in the MOT
 of $\mathrm{{}^{170}Yb}$,
 $\mathrm{{}^{174}Yb}$,
 and $\mathrm{{}^{176}Yb}$ with the same system.

The optical depth before and after the release from the MOT 
 is shown in Fig. \ref{level-mot} (b),
 which was measured by the transmission of
  the weak probe with the intensity of $0.06~\mathrm{\mu W/mm^2}$
  at the resonant frequency $\omega^{(3/2)}$.
This data can be fitted as an exponential function
 $d\exp(-t/\tau)$, where $d$ and $\tau$ are free parameters,
 and $t$ is the time of flight.
As the result, $N\sigma_0L$
 in the probe region is estimated as
 $N\sigma_0L=3d/2=7.5\times 10^{-2}$,
 and the decay time is estimated as $\tau=2.2~\mathrm{ms}$.
The decay after the release was due to the expansion
 of the atom distribution.
Since the beam waist of the probe light was $w=0.5~\mathrm{mm}$,
 the average velocity of the atom can be estimated
 as $v\sim w/\tau=0.2~\mathrm{m/s}$,
 which was three order of magnitude smaller than the atomic beam.

\begin{figure}[h]
 \begin{center}
 \includegraphics{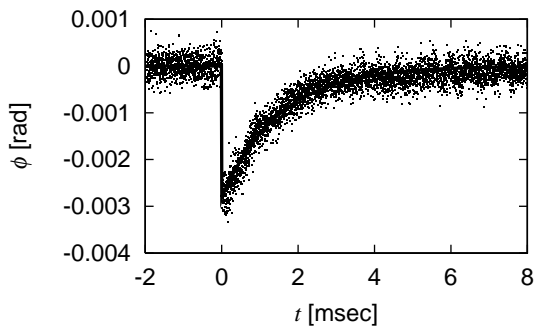}
 \caption{
   Experimental result of the rotation angle (dot).
   The rotation angle decays caused by the expansion 
   of the atom distribution from the probe region.
   The solid line is the theoretical curve estimated from
    Fig. \ref{level-mot} (b).
 }
 \label{expansion}
 \end{center}
\end{figure}

With this setup,
 we polarized the atoms via the optical pumping 
 after the release from the MOT,
 while probing the Faraday rotation with the intensity
 of $0.3~\mathrm{\mu W/mm^2}$.
The detuning of the probe beam was
 set at $\omega-\omega^{(3/2)}=2\pi\times 0.16$~GHz,
 which corresponds to the center of the two hyperfine resonances
 (see Fig. \ref{pfr-1_2} and also Fig. \ref{polspec}(a)).
At this probe frequency,
 $\phi/pN\sigma_0L$ is $3.0\times10^{-2}$.
In Fig. \ref{expansion}, we show the experimental result of the Faraday rotation
 and the fitting curve by an exponential function.
The experimental result well agrees with the theoretical estimation
 assuming the perfect polarization $p=1$.
It should be noted that the optical pumping was completed
 much rapidly compared with the expansion time.

The probed atom number was $2S=N \pi w^2 L=7\times 10^5$,
 which was one order of magnitiude smaller than
 the trapped atom number.
Therefore, the whole trapped atom could not be observed.
To improve the matching between the atomic distribution
 and the probe region,
 the compression of the MOT will be required.

\subsection{Larmor precession of trapped $\mathit{{}^{171}Yb}$}
Finally,
 we observed the Faraday rotation signal of $\mathrm{{}^{171}Yb}$
 atoms trapped in a single FORT and its
  Larmor precession due to an applied magnetic field.
By using FORT,
 the decay caused by the ballistic expansion was overcome.

\begin{figure}[h]
 \begin{center}
 \includegraphics{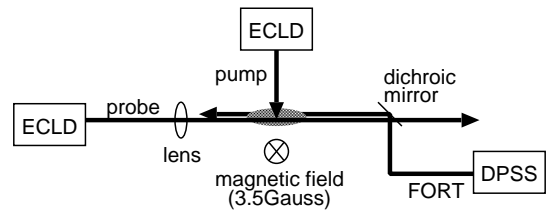}
 \caption{
  Experimental setup for the observation of the Larmor precession
  of trapped $\mathrm{{}^{171}Yb}$.
  After the optical pumping, the atom precessed
  by an applied magnetic field.
  The precession were monitored via the paramagnetic Faraday rotation
  of the probe light.
 }
 \label{fort}
 \end{center}
\end{figure}

In Fig. \ref{fort}, we describe the experimental setup.
As the FORT beam, we used a diode-pumped solid state (DPSS) laser.
After the loading of the atom from the MOT to FORT,
 we polarized the atoms via the optical pumping.
As the result, the trapped atom number was $2S=8\times 10^6$,
 the temperature was 0.1~mK,
 and the distribution was the pencil shape of
 the waist $3~\mathrm{\mu m}$ and the length $L=1~\mathrm{mm}$
 \cite{takasu03}.
The probe beam of the diameter 30~mm was
 focused by a lens of the focal length 300~mm,
 so as to match the atomic distribution.
In spite of these efforts,
 the beam waist of the probe beam at the atom distribution
 was $w=30~\mathrm{\mu m}$,
 which was not narrow enough for atomic distribution.
Therefore, effectively $N\sigma_0L$ in the expression of $\phi$ becomes smaller
 and $N\sigma_0L=2S\sigma_0/(\pi w^2)=2\times 10^2$.
The detuning of the probe beam was
 set at $\omega-\omega^{(3/2)}=2\pi\times 1.6$~GHz,
 which was the blue detuning
 so as not to induce the photoassociation \cite{takasu04}.
At this probe frequency,
 $\phi/pN\sigma_0L$ is $3.8\times10^{-4}$.
Since the magnetic field was measured as $3.5\times10^{-4}~\mathrm{T}$,
 the precession frequency was $\omega_B=2\pi\times2.6~\mathrm{kHz}$,
 derived from the gyromagnetic ratio $7.50\times10^{6}~\mathrm{Hz/T}$.

\begin{figure}[h]
 \begin{center}
 \includegraphics{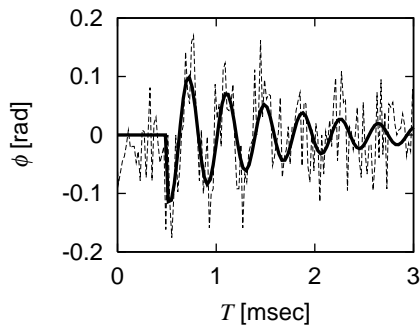}
 \caption{
  Experimental result of the rotation angle (dot line).
  The bold line is the fitting curve,
  based on the Eq .(\ref{precession-decay}).
  The rotation angle oscillates at the Larmor frequency,
  and decays by the trap loss of the atomic ensemble.
 }
 \label{precession}
 \end{center}
\end{figure}

In Fig. \ref{precession},
 we show the experimental result,
 where $T$ is the time after the optical pumping.
It should be noted that this signal represents the difference
 between the Faraday rotation signals with and without the optical pumping.
This signal can be fitted by
\begin{align}
 \phi=\Phi\exp(-T/\tau)\sin(\omega_B T+\theta),
  \label{precession-decay}
\end{align}
 where $\Phi$, $\tau$, and $\theta$ are free parameters.
The frequency agrees with the Larmor frequency $\omega_B$.
The amplitude of the rotation angle $\Phi$ also agrees with
 the assumption of perfect polarization $p=\sin(\omega_B T+\theta)$.

As the reason of the decay,
 we think that the photon scattering decreases $N\sigma_0L$.
Since the probe intensity at the atom was $I=0.70~\mathrm{mW/mm^2}$,
 the photon scattering rate becomes $r=8.7\times 10^3~\mathrm{s^{-1}}$.
The acceleration roughly becomes
 $a\sim\hbar\omega r/M_{171}c=51~\mathrm{m/s^2}$,
 where $M_{171}$ is the mass of $\mathrm{{}^{171}Yb}$.
The hold time in the trap region roughly becomes
 $\sqrt{2L/a}=6~\mathrm{ms}$.
This value is comparable to the experimental result.
As the other reason of the decay,
 the depolarization caused by the photon scattering can be included.
The trap loss and the depolarization can be suppressed
 by taking more off-resonance and reducing the intensity.
Moreover, by improving the probe light beam diameter
 so as to match the atom distribution,
 the rotation angle remains large enough
 as is calculated in Ref. \cite{takeuchi05}.

\section{Conclusions}
In this paper, we derive the theoretical value
 of the paramagnetic Faraday rotation of Yb atoms,
 and report its observation.
As the atomic samples,
 we used 
 an atomic beam,
 released atoms from a MOT,
 and trapped atoms in a FORT.
By use of the atomic beam which includes many isotopes,
 we demonstrated the polarization spectoroscpy
 by isotope-selective optical pumping.
By use of the released atom from MOT,
 we observed the Faraday rotation 
 of the ballistically expanding cold atoms.
By use of the trapped atom in FORT,
 we observed the Larmor precession by an applied magnetic field.
In these system,
 we have succeeded in the almost perfect polarization,
 which was evaluated from the rotation angle.
These results are important progresses
 for the realization of spin QND,
 spin squeezing via one-axis twisting with coherent light
  \cite{takeuchi05},
 the search of the permanent electric dipole moment
 of ytterbium atom \cite{takahashi97},
 and so on.

\section{Acknowledgements}
We thank K. Komori, Y. Iwai, and D. Komiyama
 for their experimental asisstances.
We also thank Y. Narukawa, S. Nagahama (Nichia Chemical Industries)
 and Y. Kawakami (Kyoto Univ.) for supplying the violet-LDs.
We acknowledge EpiQuest, Inc.
 for their development of the atomic oven.
This work was supported by the Strategic Information
  Communications R\&D Promotion Programme (SCOPE-S)
  and Grant-in-Aid for the 21st century COE,
  ''Center for Diversity and Universality in Physics''
  from the Ministry of Education, Culture, Sports, Science,
  and Technology (MEXT) of Japan.
M. Takeuchi and Y. Takasu are supported by JSPS.


\begin{thebibliography}{99}
\bibitem{budker02}
    See for example,
    D. Budker, W. Gawlik, D.F. Kimball, S.M. Rochester, V.V. Yashchuk, A. Weis,
    Rev. Mod. Phys. \textbf{74,} 1153 (2002).
\bibitem{happer72}
    See for example,
    W. Happer, 
    Rev. Mod. Phys. \textbf{44,} 169 (1972).
\bibitem{takahashi99}
    Y. Takahashi, \textit{et al.},
    Phys. Rev. A \textbf{60,} 4974 (1999).
\bibitem{kuzmich98}
    A. Kuzmich, L. Mandel, J.Janis, Y.E.Young, R. Ejnisman, and N.P.  Bigelow,
    Europhys. Lett. \textbf{42,} 481 (1998).
\bibitem{kuzmich99}
    A. Kuzmich, L. Mandel, J. Janis, Y.E. Young, R. Ejnisman, and N.P. Bigelow,
    Phys. Rev. A \textbf{60,} 2346 (1999).
\bibitem{kuzmich00}
    A. Kuzmich, L. Mandel, and N.P. Bigelow,
    Phys. Rev. Lett. \textbf{85,} 1594 (2000).
\bibitem{julsgaard01}
    B.Julsgaard, A.Kozhekin, and E.S. Polzik,
    Nature(London) \textbf{413,} 400 (2001).
\bibitem{geremia04}
    J.M. Geremia, J.K. Stockton, and H. Mabuchi,
    Science \textbf{304,} 270 (2004).
\bibitem{julsgaard04}
    B. Julsgarrd, J. Sherson, J.I. Cirac, J. Flur\'{a}\v{s}ek, and E.S. Polzik,
    Nature(London) \textbf{432,} 482 (2004).
\bibitem{terashima05}
    H. Terashima and M. Ueda,
    quant-ph/0507020.
\bibitem{takahashi97}
    Y. Takahashi,
    M. Fujimoto, T. Yabuzaki, A.D. Singh, M.K. Samal, and B.P. Das, 
    in \textit{Proceedings of CP Violation and its origins},
    edited by K. Hagiwara (KEK Reports, Tsukuba, 1997).
\bibitem{isayama99}
    T. Isayama, \textit{et al.},
    Phys. Rev. A \textbf{59,} 4836 (1999).
\bibitem{honda99}
    K. Honda, \textit{et al.},
    Phys. Rev. A \textbf{59,} R934 (1999).
\bibitem{kuwamoto99}
    T. Kuwamoto, \textit{et al.},
    Phys. Rev. A \textbf{60,} R745 (1999).
\bibitem{honda02}
    K. Honda, \textit{et al.},
    Phys. Rev. A \textbf{66,} 021401(R) (2002).
\bibitem{takasu03}
    Y. Takasu, \textit{et al.},
    Phys. Rev. Lett. \textbf{90,} 023003 (2003).
\bibitem{metcalf99}
    See for example, Harold J. Metcalf and Peter van der Straten,
    \textit{Laser Cooling and Trapping}
    (Springer, 1999).
\bibitem{suter97}
    See for example, Dieter Suter,
    \textit{The Physics of Laser-Atom Interactions}
    (Cambridge University Press, 1997).
\bibitem{sakurai95}
    See for example, J.J. Sakurai,
    \textit{Modern Quantum Mechanics}
    (Addison-Wesley,1995).
\bibitem{banerjee03}
    A. Banerjee, U.D. Rapol, D. Das, A. Krishna, and V. Natarajan,
    Europhys. Lett. \textbf{63,} 340 (2003).
\bibitem{komori03}
    K. Komori, \textit{et al.},
    Jpn. J. Appl. Phys. \textbf{42,} 5059 (2003).
\bibitem{takasu04}
    Y. Takasu, \textit{et al.},
    Phys. Rev. Lett. \textbf{93,} 123202 (2004).
\bibitem{takeuchi05}
    M. Takeuchi, \textit{et al.},
    Phys. Rev. Lett. \textbf{94,} 023003 (2005).
\end{thebibliography}
\end{document}